\documentstyle[preprint,aps]{revtex}

\newcommand{\bra}[1]{\mbox{$\langle #1 \mid $}}
\newcommand{\ket}[1]{\mbox{$\mid #1 \rangle $}}

\begin{document}

\draft

\title{Tunneling Gap as Evidence for Time-Reversal Symmetry Breaking
	at Surfaces of High-Temperature Superconductors}

\author{R. B. Laughlin\cite{laughlin}}
\address{Department of Physics, Stanford University, Stanford, CA 94305\\
	and \\
	Lawrence Livermore National Laboratory,
	P. O. Box 808, Livermore, CA 94550}

\maketitle

\begin{abstract}
It is argued that recent Josephson junction and point-contact tunneling
experiments, interpreted as intended by their authors, indicate that
time-reversal symmetry breaking occurs at surfaces of cuprate superconductors.
The variation among experiments and the failure of previous searches to find
$T$-violation are ascribed to disorder and effects of 3-dimensionality.  The
``anyon" approach to the $t$-$J$ model is shown to predict a conventional BCS
order parameter of $d_{x^2-y^2} + i \epsilon \ d_{xy}$ symmetry, with
$\epsilon $ roughly 3 times the doping fraction $\delta $, which is consistent
with these experiments but not demonstrated by them.
\end{abstract}


\narrowtext

The purpose of this Letter is to point out that time-reversal symmetry
breaking\cite{A}, the key prediction of the ``anyon" approach to the
high-temperature superconductivity problem, may already have been demonstrated
in a series of recent experiments conducted for other purposes.  While it is
wise to be cautious, particularly in light of previous failures to detect
$T$-violation\cite{B}
and the possibility that one or more of these experiments may
later prove to be wrong or misinterpreted, the implication of the experiments
as they now stand is clear.  The relevant experiments are the
photoemission\cite{C}
and light scattering\cite{D} measurements of the gap anisotropy, the microwave
measurements of the low-temperature conductivity\cite{E}, the Josephson phase
coherence experiments of Wollman {\it et al.}\cite{F},
and the scanning tunneling
microscope measurements of Hasegawa {\it et al.}\cite{G} reporting an intrinsic
energy gap.  The essence of the argument is that all of these except the last
point to the occurrence of a conventional BCS order parameter of $d_{x^2-y^2}$
symmetry, the behavior commonly found in theories based on incipient
antiferromagnetism of the conducting electrons\cite{H}.  This, however, is
fundamentally incompatible with the last experiment unless the order parameter
is {\it complex}, which is impossible unless the ground state violates $T$.
Thus the simultaneous occurrence of $d$-wave superconductivity and an energy
gap in any part of the sample, if true, constitutes definitive evidence for
$T$-violation.

A $d_{x^2-y^2}$ order parameter $\Delta _k$ is distinguished from a
conventional $s$-wave order parameter by sign reversals.  In high-Tc
superconductors this is most easily discussed in terms of the idealized
electron energy band

\begin{equation}
E^0_k = -2t_0\left[ \cos (k_xb) + \cos (k_yb)\right]
\end{equation}

\noindent
where $t_0$ is an energy parameter and $b$ is the bond length of a
2-dimensional square lattice.  One imagines creating a fermi sea by filling the
states with $E_k<0$.  Allowing the electrons to interact through weak
near-neighbor spin exchange\cite{H}
then leads to a superconducting state for which
the quasiparticles have energies

\begin{equation}
E_k = \pm  \sqrt{(E^0_k)^2+|\Delta_k|^2}
\end{equation}

\noindent
where

\begin{equation}
\Delta _k = \Delta ^0\left[ \cos (k_xb) - \cos (k_yb)\right].
\end{equation}

\noindent
The gap parameter $\Delta _k$ is positive in the $x$-direction, negative in the
$y$-direction, and zero at the nodes in between.  In an $s$-wave superconductor
$\Delta _k$ is the constant $\Delta ^0.$

Of the experiments listed above, only the Josephson experiment of Wollman
{\it et al.}\cite{F}
can directly sense the sign reversal of $\Delta _k$, even in
principle.  It accomplishes this by detecting electric currents spontaneously
generated in a loop of Pb wire connected between the $x-$ and $y-$faces of a
high-Tc superconductor.  Although the result reported by Wollman {\it et
al.}\cite{F} is positive and consistent with the large body of circumstantial
evidence for $d$-wave pairing, it has not yet been reproduced and is quite
controversial.  In what follows we shall assume that this experiment is right.
It must be emphasized that it is the {\it only} direct evidence we have for
sign reversal of $\Delta _k.$

This experiment is supported by considerable indirect evidence for $d$-wave
superconductivity with nodes at $k = (\pm \pi /2b,\pm \pi /2b)$, in particular
by the presence in {\it all} samples of low-energy excitations in the
superconducting state.  The evidence is, unfortunately, complicated by the
disorder effects that plague these materials.  For example, Giaver tunneling
finds states in the gap so commonly that the claim of Hasegawa {\it et.
al.}\cite{G}
to have observed a clean gap, the experiment motivating this Letter,
is widely questioned.  At the same time, the large zero-bias conductance seen
in most tunneling experiments is commonly dismissed as an artifact of disorder
at the tunnel contact\cite{I}.
The $T^2$ deviation of the penetration depth from
its zero-temperature value, and its crossover to linear$-T$ behavior above
about 5 $K$, seen in high-quality samples\cite{J} is consistent with
$d_{x^2-y^2}$ superconductivity only if disorder that is difficult or
impossible to eliminate from the samples\cite{K} is assumed to exist.  The same
is true of the zero-temperature microwave conductivity\cite{E,J}.
The intrinsic
heat capacity below 10 $K$ is not known for any high-$T_c$ superconductor
because the signal is always swamped by a large Schottkey-like heat
capacity\cite{L} similar to that expected of a spin glass.  That at least some
of these excitations are intrinsic and attributable to a $d$-wave node is
indicated by several less accurate or model-dependent experiments, the most
accessible of which is the angle-resolved photoemission work of Shen {\it et
al.}\cite{C}.
This reports measureable changes to the quasiparticle energies $E_k$
resulting from cooling the sample through its superconducting transition
{\it except} near this special value of $k$.  The complex temperature and
polarization dependence of inelastic light scattering is accounted for
quantitatively by $d_{x^2-y^2}$ superconductivity if reasonable assumptions are
made about the relevant matrix elements$^4$.  The same is true for the
extensive magnetic resonance data\cite{M}.  While the situation is still
confusing and controversial, it is clear that a large body of experimental
evidence is consistent with the simultaneous presence in all samples of both
disorder and a $d$-wave node.

Let us now consider the tunneling experiment of Hasegawa {\it et al.}\cite{G}.
It
is very important for our argument that this experiment was performed with a
scanning tunneling microscope tip, that it reported a well-developed energy gap
only at certain places on the sample, and that it was inconsistent with
conventional thin-film tunnel junction experiments, which never reveal a full
gap\cite{I}.
Barring the possibility that it is an experimental artifact, such as
coulomb blockade\cite{N} or an effect of anisotropic tunneling, this result
implies that a genuine energy gap in the quasiparticle spectrum develops at
some places on the surface.  The cleanliness of the reported gap suggests that
this gap is intrinsic.  Why it should develop only on islands is an open
question.  One possiblity is that it is destroyed almost everywhere on the
surface by disorder.  That disorder has the potential to do this, provided that
the order parameter has sign reversals, is well known\cite{K}.
This explanation
probably cannot account for the absence of the gap in the bulk, although
further experimental studies are required to tell for certain.  A more likely
explanation in this case is that the gap is suppressed because of the magnetic
fields it would create.  However, regardless of the mechanism by which the gap
is destroyed, its existence {\it anywhere} in the sample indicates
$T$-violation, since an energy gap is forbidden as a matter of principle in a
$d$-wave superconductor with a real order parameter.

Let us now consider the specific complex order parameter

\begin{equation}
\Delta ^{\rm chiral}_k = \Delta ^0 \left[  \cos (k_xb) - \cos (k_yb) +
i\epsilon \ \sin (k_xb)\sin (k_yb)\right]
\end{equation}

\noindent
where $\epsilon $ is a number, implicit in the anyon technique.  Because the
imaginary part of this order parameter has sign reversals, it is susceptible to
destruction by non-magnetic disorder\cite{K} and thus consistent with the idea
that $\epsilon $ might be zero everywhere except on islands.  That there should
be such an order parameter was first suggested by Rokhsar\cite{O}, who also
pointed out that it was potentially inconsistent with the known phase diagram
of these materials.  Since the real and imaginary parts of $\Delta ^{chiral}_k$
lie in different irreducible representations of the lattice point group, they
cannot mix in the Ginzberg-Landau functional, and thus must develop as separate
order parameters as the temperature is lowered.  This would give 2 phase
transitions, rather than the observed 1.  However, the suppression of the the
smaller $d_{xy}$ order parameter almost everywhere in the sample would resolve
this paradox by preventing the acquisition of long-range order and killing the
lower phase transition.  It would also kill all effects of global
$T$-violation, and account for the failure of previous searches to find
them\cite{B}.

Let us now show that this order parameter is implicit in the anyon approach to
the $t-J$ Hamiltonian

\begin{equation}
{\cal H}_{t-J} = \sum_{<j,k>} \ \biggl\{ -t \sum_{\sigma}
c^{\dag}_{j \sigma}c_{k \sigma} + {J
\over 2} S_j \cdot S_k
\biggr\}
\end{equation}

\noindent
and estimate its magnitude.  It should be noted such an order
parameter is not uniquely obtained this way, but also arises in certain
simple BCS Hamiltonians\cite{P}.  As usual, $S_j$ denotes the spin operator
for the $j^{\rm th}$ site, ${<j,k>}$ denotes a sum on near-neighbor pairs, with
each pair counted twice, and $J = 0.1$ eV and $t = 0.5$ eV are the
spin-exchange and electron hopping matrix elements.  We assume that the
lattice has $N$ sites and $M = N\delta $ holes.

The anyon superconducting state for the $t-J$ model is an electron
wavefunction of the form

\begin{equation}
\mid\Psi\rangle = \sum_{\ell_1,...,\ell_M} a_{\ell_1,...,\ell_M}
\mid{\ell_1,...,\ell_M }\rangle
\end{equation}

\noindent
where $\ket{\ell_1,...,\ell_M}$ is a basis wave function describing "holons" at
sites $\ell_1,...,\ell_M$.  This is given explicitly by\cite{Q}

\begin{equation}
\ket{\ell _1,...,\ell _M} =
\prod_j^N (1-n_{j\uparrow}n_{j\downarrow})
\prod_{\alpha}^M (1-n_{\ell_{\alpha}\uparrow} -n_{\ell_{\alpha}\downarrow} )
\ket{ \Psi_{\rm flux} }
\end{equation}

\noindent
where $n_{j\sigma } = c_j^{\dag} \sigma^c j \sigma$ is the number operator
for an electron of spin $\sigma $ at site $j$, and $\ket{\Psi_{flux}} $ is the
ground state of the commensurate flux Hamiltonian

\begin{equation}
{\cal H}_{\rm flux} = - t \ \sum_{<j,k>} \sum_{\sigma}
\exp{\biggl\{ i {\pi \over 2} (1 - \delta) (x_j - x_k) (y_j + y_k)
/ b^2 \biggr\}}
c^{\dag}_{j \sigma}c_{k \sigma}
\end{equation}

\noindent
with $N-M$ electrons.  ${\cal H}_{flux}$ violates $T$, as is required for the
holon basis to be defined\cite{Q}.  It is also possible to construct the basis
using ${\cal H}^*_{flux}$, in which case $T$ is violated in the opposite sense.
The holons defined by Eq. (7) obey $1/2$ fractional statistics\cite{Q} in that
varying the expansion coefficients in Eq. (6) to minimize the expected energy
$\bra{\Psi}{\cal H}_{t-J}\ket{\Psi}$ is equivalent, when $\delta $ is small, to
solving the fermion problem\cite{Q}

\begin{eqnarray}
{\cal H}_{\rm anyon} & = &
\sum_j{\hbar^2 \over {2m^*}}|P_j+{e \over c}A_j |^2 \\
A_j & = & {1 \over 2} {\hbar c \over e} \sum_{k \not= j}
{{\hat z} \times (r_j - r_k) \over {|r_j - r_k|^2}}
\end{eqnarray}

\noindent
with {\it isospin}.  The latter corresponds to the valley degeneracy of
the holon band structure, which in the $\delta \rightarrow 0$ limit is
described by Eq. (8) {\it without spin}\cite{Q}.  Since the holon dispersion
relation in this limit is given by

\begin{equation}
E^{\rm holon}_k \cong  \pm  2t \sqrt{\cos ^2(k_xb)+\cos ^2(k_yb)}\>,
\end{equation}

\noindent
we find that the valley minima occur at $k = (0,0)$ and $(\pi /b,0)$ and are
characterized by the mass $m^* \cong  {\hbar}^2/(\sqrt{2} b^2 t).$

The first step in the order parameter calculation is to compute the
{\it non-local} order parameter of the continuum anyon gas described by Eq.
(9)\cite{R}.  We will take the ground state to be

\begin{equation}
\ket{\Phi _{\rm anyon}} \cong
\prod_q \left\lbrace  {1\over 2} \exp (-|q|/\sqrt{\pi \rho }
\rho _q\rho _{-q})\right\rbrace  \ket{\Phi _{HF}}
\end{equation}

where $\rho  = \delta /b^2$ is the particle density, $\rho _q =
\Sigma_j\exp (iq\cdot r_j)$
is the density operator, and $\ket{\Phi _{HF}}$ is the
ground state of the Hamiltonian obtained by substituting the mean-field vector
potential ${<A>} = hc/(2e)\rho \ y\hat x$
for $A_j$ in Eq. (9).  The prefactor in
this expression is the usual modification of the Hartree-Fock ground state
implicit in the random phase approximation.  Let $\psi _+(z)$ and $\psi _-(z)$
denote the operators annihilating holons with ``up" and ``down" isospin,
respectively, at $z = x+iy$ in the fermi representation.  The nonlocal order
parameter is given in terms of these by\cite{R}

\begin{eqnarray}
\lefteqn{\int \bra{ \Phi_{\rm anyon} }\psi_+^{\dag}(z_1)\psi_-^{\dag}(z_1)
\biggl\{ \psi_+^{\dag} (z) \psi_+ (z) + \psi_-^{\dag} (z)
\psi_- (z) \biggr\}
\psi_+ (z_2) \psi_- (z_2) \ket{ \Phi_{\rm anyon} }} \nonumber
\\
& & \times {(z_1^* - z^*) \over {|z_1-z|}}
{(z_2 - z) \over{|z_2 - z|}} d^2z \qquad
{\cong \atop{|z_1 - z_2| \rightarrow \infty}} \qquad (1.9 \rho)^2
\end{eqnarray}

\noindent
The numerical value of 1.9 $\rho $ reported here for the first time is obtained
using the hypernetted chain technique\cite{Q}.  We will adopt the notation
$\langle\psi _+(z)\psi _-(z)\rangle = 1.9\ \rho $ as shorthand for this result.

The second step is to convert this continuum order parameter to the site basis.
The unitary transformation relating $\psi _+(z)$ and $\psi _-(z)$ to $\psi
(j)$, the fermi operator to annihilate a holon at site $r_j = (\ell _j,m_j) b$,
is simply the the matrix of 1-body eigenstates of Eq. (8) at the two valley
minima.  We thus have

\begin{equation}
\psi (j) \cong
\sqrt{2} b \left\lbrace \cos \left[ ({1\over 8} - {1\over 2}m^2_j)\pi \right]
\psi_+ (z) + (-1)^{\ell _j} \sin \left[ ({1\over 8} +
{1\over 2}m^2_j)\pi \right]  \psi _-(z)\right\rbrace
\end{equation}

\noindent
In obtaining this expression, we have imagined the sample to be divided into
4-site cells, and that $z$ defines the center of the cell containing $j$.
Then, substituting Eq. (13) into Eq. (12), and using the fact that $\psi _+(z)$
and $\psi _-(z)$ anticommute with themselves and each other, we obtain

\begin{eqnarray}
\lefteqn{\langle 0 \mid \Psi(1) \Psi(2) \mid 0 \rangle \cong
1.9 \delta (-1)^{\ell_1 + m_2 + (\ell_1 + \ell_2)m_1}} \nonumber \\
& & \times \frac{1}{\sqrt{2}}\left(
\left[ 1-(-1)^{\ell_1 + \ell_2 + m_1 + m_2}\right] - \left[
1-(-1)^{(\ell_1 + \ell_2)(m_1 + m_2)} \right] \right)
\end{eqnarray}

\noindent
The third step is to evaluate the matrix element of
$c_{j \uparrow}^{\dag} c_{k \downarrow}^{\dag}$ to annihilate a holon pair into
the vacuum.  This matter has been studied
extensively in previous papers\cite{Q}
and is too involved to discuss in detail here.  We shall simply quote the
result

\begin{eqnarray}
\lefteqn{\bra{\ell '_1,...,\ell '_M}
c_{k'\downarrow }c_{j'\uparrow }
c_{j \uparrow}^{\dag} c_{k \downarrow}^{\dag}\ket{
\ell _1,...,\ell_M} \cong} \nonumber \\
& &  \delta _{\ell _1j} \delta _{\ell _2k} \delta _{\ell '_1j'}
  \delta _{\ell '_2k'}\prod_{\alpha = 3}^M
  \delta _{\ell _\alpha \ell '_\alpha }
  {({z'}_j^*+{z'}_k^*)/2-z^*_\alpha \over |(z'_j +z'_k )/2-z_\alpha |}
  {(z_j+z_k)/2-z_\alpha \over |(z_j+z_k)/2-z_\alpha |} \nonumber \\
& &  \times  \bra{j',k'}c_{j'\uparrow }c_{k'\downarrow }\ket{0} \bra{0
}c_{j \uparrow}^{\dag} c_{k \downarrow}^{\dag} \ket{
j,k } + {\rm cyclic \; permutations}
\end{eqnarray}

\noindent
where $\ket{0}$ denotes the state with no holons and $z_j=(\ell_j + im_j) b$.
The product on $\alpha $ in this expression is the same ``unwinding" factor
appearing in Eq. (12) and is the microscopic justification for its inclusion in
Eq. (12).  The remaining factor is given approximately by

\begin{eqnarray}
\bra{0 } c_{1 \uparrow}^{\dag} c_{2 \downarrow}^{\dag} \ket{ j,k } & \cong &
{(z_1 - z_2) \over {|z_1 - z_2|}} i^{(\ell_1 - \ell_2)(m_1 + m_2)}
(-1)^{\ell_1 + m_1} \nonumber \\
& \times &
\left[{\begin{array}{cl}
 (1-\delta)\exp{\lbrace -\pi /4(1-\delta)\rbrace} &
 {\rm ;\; 1\; and\; 2\; near\; neighbors} \\
 \delta \exp{\lbrace - \pi \delta \rbrace } &
 {\rm ;\; 1 \;and\; 2\; second\; neighbors} \\
 0 &
 {\rm ;\; otherwise}
\end{array}}\right]
\end{eqnarray}

The final step is to combine Eq. (15) and (16) with Eq. (14) to obtain an
expression for
$\langle c_{j\uparrow }^{\dag} c_{k\downarrow}^{\dag} \rangle$.
This is accomplished by
multiplying together the right sides of Eqs. (14) and (16) and
then dividing out the factor $(z_1-z_2)/|z_1-z_2|$.  This latter step accounts
for the transformation of the basis functions defined by Eq. (7), which are
{\it symmetric} under interchange of the holon positions, to their fermi
representation.  We obtain finally

\begin{eqnarray}
\langle c_{1 \uparrow}^{\dag} c_{2 \downarrow}^{\dag}\rangle & \cong &
i^{(\ell_1 - \ell_2)(m_2 - m_1)}
(-1)^{m_1 - m_2} \nonumber \\
 & \times & 1.9 \delta
\left[{\begin{array}{cl}
 \sqrt{2}(1-\delta)\exp{\lbrace -\pi /4(1-\delta)\rbrace} &
 {\rm ; \;1\; and\; 2\; near\; neighbors} \\
 -2\delta \exp{\lbrace - \pi \delta \rbrace } &
 {\rm ;\; 1\; and\; 2\; second\; neighbors} \\
 0 &
 {\rm ;\; otherwise}
\end{array}}\right]
\end{eqnarray}

\noindent
This is equivalent to Eq. (4) with

\begin{equation}
\epsilon  = {\sqrt{8}\delta \over 1-\delta } e^{(1- 5\delta )\pi /4}
\end{equation}

Let us now make some comments about this result.  The first is that it is quite
crude and should be compared with experiment carefully.  For example, it makes
the unphysical prediction that superconductivity occurs in the $t$-$J$ model at
any value of $J$, $t$, and $\delta $.  A more formal development of the anyon
approach using gauge theory techniques\cite{S}
cures problems of this kind, but
is too technical to discuss here.  Let us simply state the main results.  i.)
For realistic values of $J/t$ the order parameter is significantly smaller than
Eq. (17) predicts, due to retardation effects.  ii.) The equations have an
antiferromagentic instability for $\delta  \leq  .05$ that depends weakly on
$J/t$.  iii.) The commensurate flux band structure of Eq. (6) has a gap
collapse at $\delta  \cong  1/3$ which causes calculations for doping fractions
larger than this to be unreliable.  That the calculation finds conventional
superconducting order and $d$-wave symmetry is not surprising.  The anyon
approach is a legitimate variational technique for the $t$-$J$ model, which is
known by more reliable methods to have a tendency to $d$-wave pairing\cite{T}.
Also, it has been known for several years that ``flux" vacua are fundamentally
related to $d$-wave superconducting states\cite{U}.  The calculation is
significant mainly because it predicts that superconducting pairing by means of
``spin fluctuations" tends naturally to an order parameter with $d_{x^2-y^2} +
i\epsilon \ d_{xy}$ symmetry, with $\epsilon $ significantly large.

I wish to thank M. R. Beasley, A. Kapitulnik, T. H. Geballe, S. Doniach, Z.
Zou, A. A. Abrikosov, and D. J. Scalapino for numerous helpful discussions.
The work was supported primarily by the National Science Foundation under Grant
No. DMR-88-16217.  Additional support was provided by the NSF MRL Program
through the Center for Materials Research at Stanford University.


\begin{references}

\bibitem[*]{laughlin}
e--mail:  rbl@large.stanford.edu

\bibitem{A}
X.-G. Wen, F. Wilczek, and A. Zee, Phys. Rev. {\bf B39}, 11413 (1989).

\bibitem{B}
S. Spielman {\it et  al}, Phys. Rev. Lett. {\bf 69}, 123 (1990); T.W.
Lawrence, A Sz\"oke, and R. B. Laughlin, Phys. Rev. Lett. {\bf 69}, 439 (1992).

\bibitem{C}
Z.-X. Shen {\it et al}, Phys. Rev. Lett. {\bf 70}, 1553 (1993); {\it ibid}.,
J. Phys. Chem. Solids {\bf 53}, 1583 (1992); D. S. Dessau {\it et al, ibid}.
{\bf 52}, 1401 (1991); B. O. Wells {\it et al}, Phys. Rev. {\bf B46}, 11830
(1992).

\bibitem{D}
T. P. Devereaux {\it et al}, Phys. Rev. Lett. {\bf 72}, 396 (1994); T.
Staufer, {\it et al}, Phys. Rev. Lett. {\bf 68}, 1069 (1992).

\bibitem{E}
D. A. Bonn, {\it et al}, Phys. Rev. {\bf B47}, 11314 (1993).

\bibitem{F}
D. A. Wollman {\it et al}, Phys. Rev. Lett. {\bf 72}, 2134 (1993).

\bibitem{G}
T. Hasegawa {\it et al}, J. Phys. Chem Solids {\bf 54}, 1351 (1993).

\bibitem{H}
S. R. White {\it et al}., Phys. Rev. {\bf B39}, 839 (1989); P. Monthoux and
D. Pines, Phys. Rev. Lett. {\bf 69}, 961 (1992).

\bibitem{I}
J. M. Valles, Jr. {\it et al}., Phys. Rev. {\bf B44}, 11986 (1991); D.
Mandrus {\it et al}, Nature {\bf 351}, 460 (1991).

\bibitem{J}
Z. Ma {\it et al}., Phys. Rev. Lett. {\bf 71}, 781 (1993); W. Hardy {\it et
al}., Phys. Rev. Lett. {\bf 70}, 3999 (1993); S. Anlage {\it et al}. Phys. Rev.
{\bf B44}, 9764 (1991).

\bibitem{K}
M. Prohammer and J.P. Carbotte, Phys. Rev. {\bf B43}, 5370 (1991); J.
Annett, N. Goldenfeld, and S. R. Renn, Phys. Rev. {\bf B43}, 2778 (1991); F.
Gross {\it et al}. Z. Phys. {\bf B64}, 175 (1986); R. J. Radtke, K. Levin,
H.-B. Sch\"uttler, and M. R. Norman, Phys. Rev. {\bf B48}, 653 (1993);  P. J.
Hirschfeld, W. O. Putikka, and D. J. Scalapino, Phys. Rev. Lett. {\bf 71}, 3705
(1993).

\bibitem{L}
N. E. Phillips, R. A. Fisher, and J. E. Gordon, Prog. Low Temp. Phys.
{\bf 13}, 267 (1992).

\bibitem{M}
N. Bulut and D. J. Scalapino, Phys. Rev. Lett. {\bf 68}, 706 (1992); J. A.
Martindale {\it et al}., {\it ibid}. {\bf 68}, 702 (1992).

\bibitem{N}
P. A. Lee, Phys. Rev. Lett. {\bf 71}, 1887 (1993).

\bibitem{O}
D. S. Rokhsar, Phys. Rev. Lett. {\bf 70}, 493 (1993).

\bibitem{P}
F. Wenger and S. \"Ostlund, Phys. Rev. {\bf B47}, 5977 (1993); F. Wenger,
Licentiate Thesis, U. of Gothenburg, 1993).

\bibitem{Q}
Z. Zou, J. L. Levy, and R. B. Laughlin, Phys. Rev. {\bf B45}, 993 (1992);
Z. Zou and R. B. Laughlin, Phys. Rev. {\bf B42}, 4073 (1990); R. B. Laughlin
and Z. Zou, Phys. Rev. {\bf B41}, 664 (1989) ; P. B\'eran and R. B. Laughlin,
Phys. Rev. {\bf B48}, 10382 (1993).

\bibitem{R}
S. M. Girvin {\it et al}., Phys. Rev. Lett. {\bf 65}, 1671 (1990).

\bibitem{S}
A. M. Tikofsky, R. B. Laughlin, and Z. Zou, Phys. Rev. Lett. {\bf 69}, 3670
(1992).

\bibitem{T}
E. Dagotto and J. Riera, Phys. Rev. Lett. {\bf 70}, 682 (1993).

\bibitem{U}
F. C. Zhang, C. Gros, T. M. Rice, and H. Shiba, Supercond. Sci. Technol.
{\bf 1}, 36 (1988).

\end{references}
\end{document}